\documentclass[sigplan,screen]{acmart}

\newcommand{\ie}{\emph{i.e.}}
\newcommand{\eg}{\emph{e.g.}}
\newcommand{\nrexamples}{three}

\newcommand{\bugsymbol}[0]{\includegraphics[width=0.3cm]{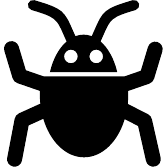}}

\usepackage{listings}
\usepackage{xcolor}
 \usepackage{microtype}
\AtBeginDocument{%
  \providecommand\BibTeX{{%
    \normalfont B\kern-0.5em{\scshape i\kern-0.25em b}\kern-0.8em\TeX}}}

\setcopyright{rightsretained}
\acmPrice{}
\acmDOI{10.1145/3563835.3567662}
\acmYear{2022}
\copyrightyear{2022}
\acmSubmissionID{onward22papers-p72-p}
\acmISBN{978-1-4503-9909-8/22/12}
\acmConference[Onward! '22]{Proceedings of the 2022 ACM SIGPLAN International Symposium on New Ideas, New Paradigms, and Reflections on Programming and Software}{December 8--10, 2022}{Auckland, New Zealand}
\acmBooktitle{Proceedings of the 2022 ACM SIGPLAN International Symposium on New Ideas, New Paradigms, and Reflections on Programming and Software (Onward! '22), December 8--10, 2022, Auckland, New Zealand}
\received{2022-07-12}
\received[accepted]{2022-10-02}

\begin{document}

\title{Intramorphic Testing} 
\subtitle{\changed{A New Approach} to the Test Oracle Problem}

\author{Manuel Rigger}
\email{rigger@nus.edu.sg}
\affiliation{%
  \institution{National University of Singapore}
  \department{School of Computing}
  \country{Singapore}
}

\author{Zhendong Su}
\email{zhendong.su@inf.ethz.ch}
\affiliation{%
  \institution{ETH Zurich}
  \department{Department of Computer Science}
  \country{Switzerland}
}

\newcommand\aname[0]{Intramorphic Testing}
\setlength\parindent{0pt}
\newcommand{\nparindent}{\hspace{10pt}}

\newcommand\changed[1]{#1}

\begin{abstract}
A \emph{test oracle} determines whether a system behaves correctly for a given input.
Automatic testing techniques rely on an automated test oracle to test the system without user interaction.
Important families of automated test oracles include \emph{Differential Testing} and \emph{Metamorphic Testing}, which are both black-box approaches; that is, they provide a test oracle that is oblivious to the system's internals.
In this work, we propose \emph{\aname{}} \changed{as a white-box methodology} to tackle the test oracle problem. 
To realize an \aname{} approach, a modified version of the system is created, for which, given a single input, a test oracle can be provided that relates the output of the original and modified systems.
As a concrete example, by replacing a greater-equals operator in the implementation of a sorting algorithm with smaller-equals, it would be expected that the output of the modified implementation is the reverse output of the original implementation.
In this paper, we introduce the methodology and illustrate it via a set of use cases.
\end{abstract}



\maketitle

\section{Introduction}

The \emph{test oracle problem} is one of the greatest challenges for software testing~\cite{Barr2015}.
A test oracle is a mechanism to check the correctness of a system's output for a set of inputs~\cite{Howden1978}.
Given that software constantly evolves and typically lacks a formal specification of the expected behavior, general test oracles are difficult to obtain.
However, \emph{partial test oracles} are still useful, as they can validate the output for some inputs~\cite{Barr2015}.
In the most straightforward case, partial test oracles are specified in the form of regression tests, where developers specify the expected output of a test case.

\nparindent{}
A number of approaches have been proposed to alleviate the test oracle problem through partial test oracles that can be applied in an automated setting (\eg{}, for automatically generated tests)~\cite{Barr2015, Pezze2014, Segura2016, Chen2018}.
The most influential ones are \emph{differential testing}~\cite{McKeeman1998} and \emph{metamorphic testing}~\cite{Chen1998}, which, similar to regression testing, approach the problem in a black-box manner; that is, they do not require access to the program's source code or internals.
Differential testing compares the output of various systems that implement the same semantics; a mismatch between outputs for the same test case indicates that at least one system is affected by a bug.
Metamorphic testing refers to a technique where, based on an existing input to a system and its output, a new input can be created for which the expected output is known.

\nparindent{}
\changed{This work presents \emph{\aname{}} as a general methodology toward the test oracle problem to complement differential testing and metamorphic testing.}
The core idea of \aname{} is to modify one or multiple components of the system under test (SUT) in a way so that the relationships between the outputs of the modified and original systems for a set of inputs are known.
Thus, different from differential testing and metamorphic testing, \aname{} is a white-box approach that assumes access to and knowledge of the system's internals.
Accordingly, we expect concrete techniques to be realized by developers---or automatically derived---rather than implemented by testers.
In this paper, we present the general idea of \aname{} and illustrate it with several concrete examples.

\nparindent{}
We believe that \aname{} techniques are already being realized and used by developers as part of an effort to create testable code.
However, they might have been viewed as an undocumented implementation detail of a test suite, rather than an instance of a more broad testing methodology.
This paper aims to address this by unifying such existing and future techniques under a common name and abstract framework, thus fueling exchange and development of \aname{} techniques.

\nparindent{}
In summary, this paper contributes the following:
\begin{itemize}
	\item \aname{}, a general conceptual white-box approach to tackling the test oracle problem;
	\item a conceptual comparison with regression testing, differential testing, and metamorphic testing;
	\item examples that illustrate the idea.
\end{itemize}

\section{Background and Motivation}

\paragraph{Test oracles}
To the best of our knowledge, the term \emph{test oracle} was coined by Howden in 1978~\cite{Howden1978}.
Since then, a number of approaches to tackle the problem have been proposed, which were summarized in surveys by, for example, Barr et al.~\cite{Barr2015} or Pezzè et al.~\cite{Pezze2014}.
Metamorphic testing was proposed by Chen et al. in a technical report in 1998~\cite{Chen1998}.
Various concrete metamorphic testing techniques were proposed that were subsequently surveyed by, for example, Segura et al.~\cite{Segura2016} or Chen et al.~\cite{Chen2018}.

\paragraph{Terminology}
Originally, a test oracle was defined to validate a system's output for a set of inputs~\cite{Howden1978}.
This view is restrictive, given that an input to the program might include changes to the device or environment.
Similarly, rather than a directly-observable program output, non-functional observations include the program's performance or changes to the device's state.
Thus, Barr et al.~\cite{Barr2015} used \emph{stimuli} for inputs and \emph{observations} for outputs to account for various testing scenarios.
We continue to use the original terminology of inputs (denoted as $I$) and outputs (denoted as $O$), but refer to them in the general sense of stimuli and observations.
We will denote the program under test as $P$.

\begin{figure*}[ht]
    \includegraphics[width=\textwidth]{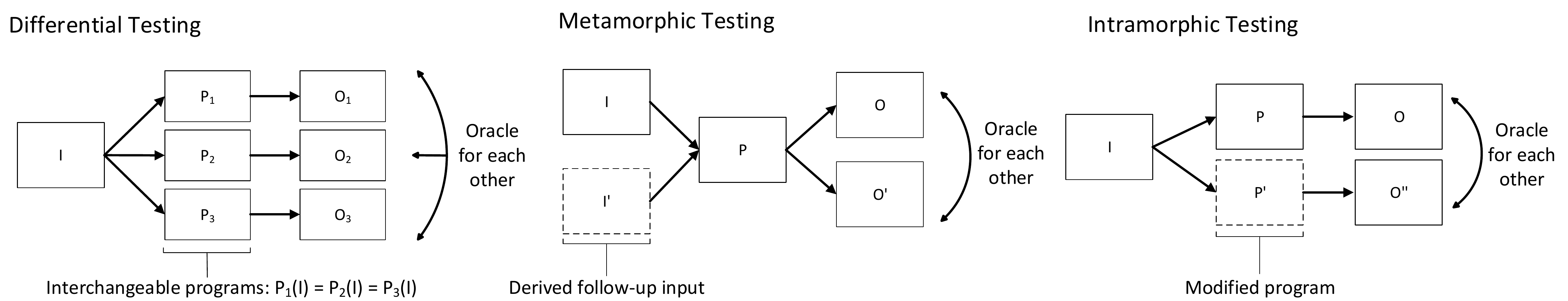}
    \caption[]{Differential testing, metamorphic testing, and \aname{} in comparison.}
    \label{fig:approaches}
\end{figure*}

\paragraph{Motivating Example}
To outline the existing techniques and our idea, let us assume a specific use case, namely that we want to test the implementation of one or multiple sorting algorithms.\footnote{A Jupyter Notebook with the code examples presented in this paper is available at \url{https://doi.org/10.5281/zenodo.7229326}.}
Let us assume that we implemented multiple sorting algorithms such as \lstinline{bubble_sort()}, \lstinline[prebreak={}]{insertion_sort()}, and \lstinline{merge_sort()}.
Let us also assume that we made a mistake when implementing the in-place \lstinline{bubble_sort()} algorithm, as illustrated in Listing~\ref{lst:buggybubblesort}; the last array index in the code listing should be \lstinline{j}, rather than \lstinline{i}.
In the subsequent paragraphs, we discuss how both instantiations of existing techniques as well as an instantiation of the proposed \aname{} technique could find the bug.
In practice, we expect that \aname{} will be realized that can find bugs that are overlooked, or difficult to find, by other testing approaches.

\begin{figure}
\begin{lstlisting}[caption={A Python implementation of \lstinline{bubble_sort} affected by a bug when swapping array elements.}, label=lst:buggybubblesort,escapeinside=||]
def bubble_sort(arr):
  length = len(arr)
  for i in range(length):
    for j in range(0, length - i - 1):
      if arr[j] > arr[j+1]:
        arr[j], arr[j+1] = arr[j+1], arr[|\textbf{\color{red}{i}}|]|\bugsymbol|
  return arr
\end{lstlisting}
\end{figure}

\paragraph{Regression Testing}
Regression testing aims to ensure that changes do not introduce bugs into the program through manually written tests in which the developer specifies the expected output.
\changed{One common way of implementing regression tests is by implementing \emph{unit tests}, where a specific unit is tested in isolation.}

\nparindent{}
To test \lstinline{bubble_sort()} and the other sorting algorithms, we could introduce \changed{unit tests} with both typical inputs as well as boundary values. 
Listing~\ref{lst:regressiontest} shows a test case that triggers the bug; sorting an array \lstinline{[3, 1, 2]} incorrectly results in \lstinline{[1, 2, 1]}, which does not match the expected array \lstinline{[1, 2, 3]}, thus revealing the bug.
Note that the test does not assume access to the system's internals; \changed{unit testing} is a black-box approach that could also be applied without access to the source code.
While \changed{unit testing} is effective and widely used, tests are typically implemented manually, and the developer needs to specify the expected outcome of the test case.

\begin{figure}
\begin{lstlisting}[caption={A manually-written \changed{unit test}.}, label=lst:regressiontest]
arr = [3, 1, 2]
bubble_sort(arr)
assert arr == [1, 2, 3]  # AssertionError, actual: [1, 2, 1]
\end{lstlisting}
\end{figure}

\paragraph{Differential Testing}
Differential testing validates a set of systems that implement the same semantics, by comparing their output for a given input.
As illustrated in Figure~\ref{fig:approaches}, given input $I$ and equivalent systems $P_1$, $P_2$, \ldots{}, $P_n$, differential testing validates that $\forall i, j: P_i(I) =P_j(I)$.
Differential testing has been applied to a variety of domains, such as, C/C++ compilers~\cite{Yang2011}, Java Virtual Machines (JVMs)~\cite{Chen2016,Chen2019}, database engines~\cite{Slutz1998}, debuggers~\cite{Lehmann2018}, code coverage tools~\cite{Yang2019}, symbolic execution engines~\cite{Kapus2017}, SMT solvers~\cite{Winterer2020}, and Object-Relational Mapping Systems (ORMs)~\cite{Sotiropoulos2021}.

\nparindent{}
As illustrated by Listing~\ref{lst:differentialtest}, we can apply differential testing by comparing the sorted arrays for multiple sorting algorithms for the same input array.
Given that the test oracle requires no human in the loop, it can be effectively paired with automated test generation; for example, in the listing, we generate random arrays as test input.
Note that the loop does not terminate; in practice, it would be reasonable to set a timeout or run the tests for a fixed number of iterations.
For an input array like \lstinline{[3, 1, 2]}, differential testing reveals a discrepancy between the output of the sorting algorithms, demonstrating the bug.
Similar to regression testing, differential testing is a black-box approach; for example, we could have also compared sorting algorithms implemented in various languages based on checking their output alone.

\begin{figure}
\begin{lstlisting}[caption={Differential testing using multiple implementations of sorting algorithms.}, label=lst:differentialtest,escapeinside=||]
sorting_algorithms = [bubble_sort, merge_sort, insertion_sort]
while True:
  arr = get_random_array() # e.g., [3, 1, 2]
  sorted_arrays = [alg(arr.copy()) for alg in sorting_algorithms]
  all_same = all(sorted_arr == sorted_arrays[0] for sorted_arr in sorted_arrays) 
  assert all_same
\end{lstlisting}
\end{figure}

\paragraph{Metamorphic Testing}
Metamorphic testing uses an input to a system and its output to derive a new input for which a test oracle can be provided via so-called \emph{Metamorphic Relations (MRs)}.
This is illustrated in Figure~\ref{fig:approaches}.
Given an input $I$ and $P(I)=O$, a follow-up input $I'$ is derived, so that a known relationship between $O$ and $P(I')=O'$ is validated.
Metamorphic testing is a high-level concept and finding effective MRs is often challenging; MRs for testing various systems such as compilers~\cite{Le2014}, database engines~\cite{Rigger2020b,Rigger2020}, SMT solvers~\cite{Winterer2020b}, Android apps~\cite{Su2021}, as well as object detection systems~\cite{Shuai2020} have been proposed in the literature.

\nparindent{}
As Listing~\ref{lst:metamorphictest} shows, we designed a MR that checks whether the relative order of sorted elements is maintained when an element is removed from an input array.
For example, given an input array $i1=[3, 1, 2]$ and a correctly-sorted array $o1=[1, 2, 3]$, we derive a new input by removing one element $e$ from the input array, for example, $e=2$, resulting in a new input array $i2=[3, 1]$, for which we can infer the expected result by removing $e$ from $o1$, that is, $o2=[1, 3]$.
This specific idea enables finding the bug as well.
When passing $[3, 1, 2]$ as input array, the incorrect output $[1, 2, 1]$ is produced; when passing $[3, 1]$ as input by removing $2$, the output is $[1, 3]$, rather than $[1, 1]$, breaking the MR's assumption.
As with regression testing and \changed{unit testing}, metamorphic testing is a black-box approach.
In contrast to differential testing, a single implementation of a system (\eg{}, sorting algorithm) is sufficient to realize the technique.

\begin{figure}
\begin{lstlisting}[caption={Metamorphic testing by comparing whether the relative order is maintained for a smaller array.}, label=lst:metamorphictest,escapeinside=||]
while True:
  arr = get_random_array()
  if len(arr) >= 1:
    sorted_arr = bubble_sort(arr.copy())
    random_elem = random.choice(sorted_arr)
    arr.remove(random_elem)
    sorted_smaller_arr = bubble_sort(arr)
    sorted_arr.remove(random_elem)
    assert sorted_arr == sorted_smaller_arr
\end{lstlisting}
\end{figure}

\paragraph{\aname{}}
\sloppy
In this work, we propose \aname{} to tackle the test oracle problem by changing the system under test so that, for a given input and the original system's output, an oracle for the output of the changed system can be derived.
Figure~\ref{fig:approaches} illustrates the approach.
Given a program $P$, a new program $P'$ is derived for which, given an input $I$, a known relationship between the two program's outputs (\ie{}, $O=P(I)$ and $O'=P'(I)$) is validated.
Similar to metamorphic testing, \aname{} is a high-level idea and conceptualization, for which many instantiations are possible.

\nparindent{}
To realize an \aname{} technique for our use case, we could implement another alternative sorting implementation \lstinline{bubble_sort_reverse()} as a potential replacement for \lstinline{bubble_sort()} that sorts the array in descending order.
Thus, the expectation that we could check is that, by reversing one of the two output arrays, two equivalent arrays are obtained (see Listing~\ref{lst:changetest}).
For example, for an input \lstinline{[3, 1, 2]}, we would expect an output \lstinline{[3, 2, 1]} for \lstinline{bubble_sort_reverse()} rather than \lstinline{[1, 2, 3]}.
This concrete \aname{} realization also detects the bug.
Even if \lstinline{bubble_sort_reverse()} is affected by the same index bug (\ie{}, by replacing only the comparison operator in Listing~\ref{lst:buggybubblesort}), it would detect the bug, since \lstinline{bubble_sort()} returns \lstinline{[3, 2, 3]} as an output and \lstinline{bubble_sort_reverse()} would return \lstinline{[1, 2, 1]}.
Note that, in order to realize the approach, we modified the system under test by adding a new function, meaning that the technique is a white-box approach; alternatively, we could have also added an additional function argument to define the sort order, or changed the function directly to manually test the assumption underlying the test oracle.
The modification of the original program discriminates this technique from regression testing, differential testing, and metamorphic testing, which are all black-box techniques.

\begin{figure}
\begin{lstlisting}[caption={An \aname{} realization that adds an additional reverse sorting function, whose reversed output array is compared with the sorted output array of the original sorting function.}, label=lst:changetest,escapeinside=||]
while True:
  arr = get_random_array()
  sorted_arr = bubble_sort(arr.copy())
  reverse_sorted_arr = bubble_sort_reverse(arr.copy())
  sorted_arr.reverse()
  assert sorted_arr.reverse() == reverse_sorted_arr
\end{lstlisting}
\end{figure}

\section{\aname{}}

In this section, we present \aname{} and its scope as well as its assumptions.

\begin{figure}[tb]
    \includegraphics[width=\columnwidth]{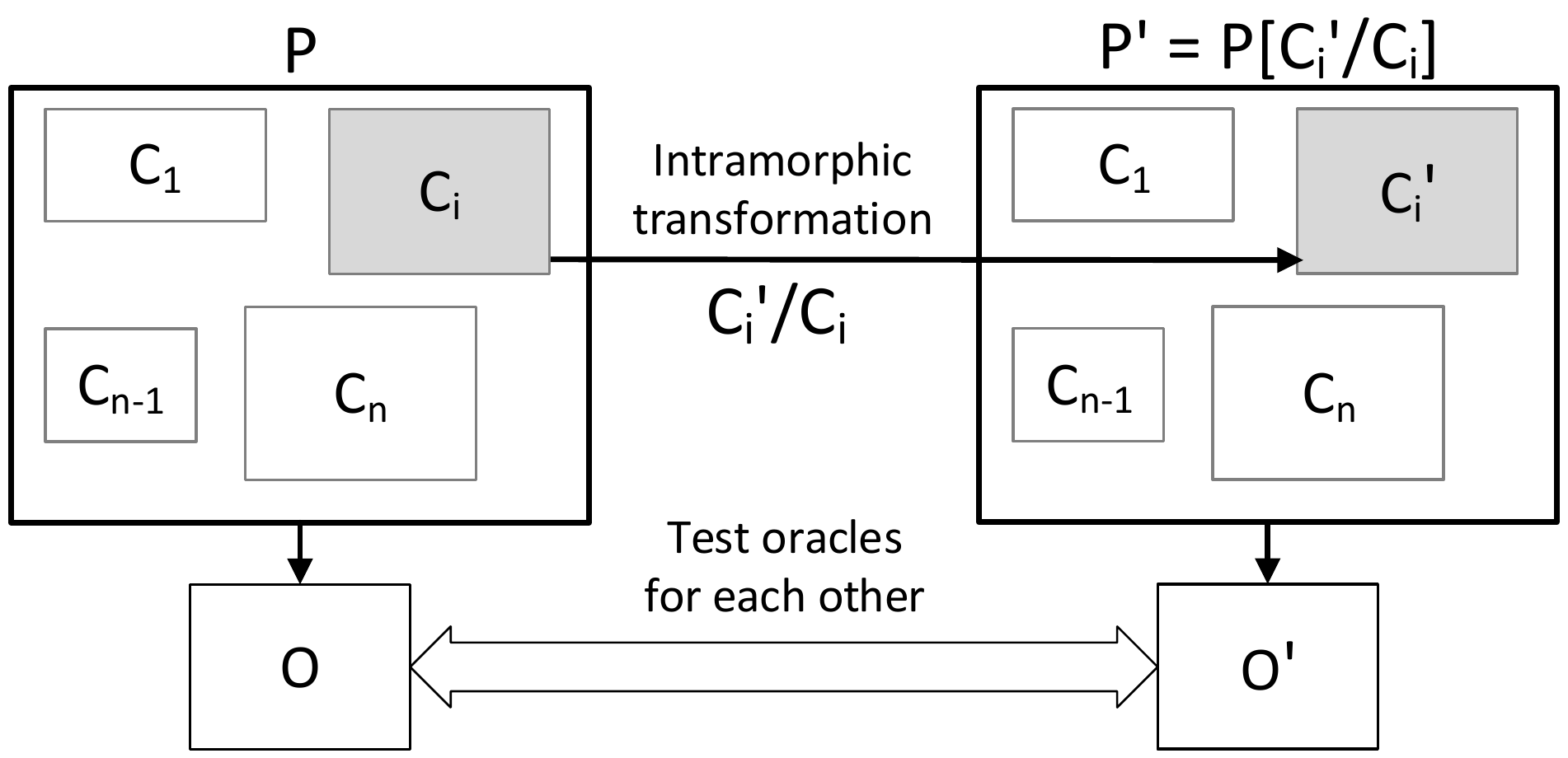}
    \caption[]{The core idea of \aname{} is to replace a component so that the output is influenced in a known way and can be related to the original system's output.}
    \label{fig:intramorphic}
\end{figure}

\paragraph{Programs as a composition of components}
It is intuitive to think of a program $P$ as a composition of components that work together to achieve a certain desired functionality.
For example, graphical interfaces often implement the \emph{model-view-controller} pattern where the program consists of three high-level components; the model represents the application's data structure, the view its (visual) representation, and the controller accepts input and converts it to commands for the model of the view~\cite{Gamma1995}.
Each such component in turn likely consists of individual components, which can be modules or classes, blocks of code, operators, or expressions---we do not prescribe any particular granularity to components.
Based on this understanding, we can view $P$ as a function $P(C_1, \ldots{},C_k)$, where $C_1,\ldots{},C_k$ are $P$'s components.

\paragraph{\aname{}}
Figure~\ref{fig:intramorphic} illustrates the idea of \aname{}.
Programmers developing $P$ are expected to have an intuition or concrete understanding of how changing a component $C_i$ to $C_i'$ affects the overall program.
Let us assume an input $I$ to $P$, such that $P(I) = O$, that is, $O$ is the output from running the program $P$ on input $I$.
Let $C_i$ be a component of $P$, and $C_i'$ be a modified component derived from $C_i$.
Let $P’ = P[C_i’/C_i]$, that is, $P'$ is the program where the component $C_i$ is replaced by the component $C_i'$.
We refer to the method used to replace the component as an \emph{intramorphic transformation}.
This local change induces a global expectation at the program level.
That is, we anticipate $P'(I)$ to change in a certain way with respect to $P(I)$.
We refer to this expectation as an \emph{intramorphic relation} and can validate it on the outputs of $P'(I)$ and $P(I)$ by running the programs; if the expectation is not met, program $P$ or $P'$ is affected by a bug.

\paragraph{Challenges}
Designing \aname{} techniques is challenging.
Given that the main goal of testing is to uncover bugs, an ideal transformation should be effective and yield as few false alarms as possible---ideally none.
Considering that developers need to implement the \aname{} technique, doing so should require as little effort as possible.
Conceptually, $P$ and $P'$ are separate programs. However, in practice, the approach needs to be integrated into the developers' workflows, where maintaining two separate program versions seems impractical---principled approaches for maintaining the program variants are needed.
While we present examples where we addressed these challenges, we expect that future research will address them for various specific domains and use cases.

\paragraph{Special case P(I) = P(I')}
A special case of \aname{} is when $P(I)$ = $P(I')$, that is, both the original and modified programs produce the same output, which can be achieved using various ways.
This can be due to a semantics-preserving transformation on $P$, producing an equivalent program variant.
Alternatively, multiple components might be available that provide the same functionality; for example, $C$ might be a bubblesort, while $C'$ a quicksort, meaning that they can be used interchangeably for most purposes.
It is also plausible that on the source code level, $P=P'$, that is, the two program versions are equivalent, but that, when compiled to machine code, $P\neq{}P'$, that is, the versions differ.
For example, different binary versions could be obtained by compiling $P$ and $P'$ with different compilers, different optimization levels, or different static application options (\eg{}, using macro metaprogramming~\cite{Liebig2010}), which closely relates to differential testing.
Conceptually, this special case also relates to N-version programming~\cite{Chen1978}, where multiple programs are developed based on the same initial specification.

\paragraph{Classification}
\aname{} can be classified along various dimensions:
\begin{itemize}
	\item the granularity of the replaced component $C_i$ (\eg{}, reaching from a replaced operator to a replaced system in a system of systems);
	\item the format of the program $P$ (\eg{}, source code or binary code);
	\item how the metamorphic transformation is applied (\eg{}, by adding a new source code component or directly replacing it);
	\item the degree of automation for the intramorphic transformation (\eg{}, whether the transformation is applied manually or can be automatically applied to many components);
	\item whether the intramorphic relation is complete (\ie{}, whether the expected output can be given for any input);
	\item whether the approach can result in false alarms---in general, it is desirable for an automated testing approach that it only detects real bugs.
\end{itemize}

\paragraph{White-box approach}
\aname{} is a white-box approach, since it relies on modifying $P$.
This contrasts the approach from differential testing and metamorphic testing, which are both black-box approaches.
This influences the target audience of the testing approach; \aname{} might be primarily applicable for developers who want to test their system, as they have a concrete understanding of the system that they are developing.
Similarly to approaches for finding metamorphic relations~\cite{Kanewala2013,Zhang2014}, future approaches to identifying and applying intramorphic transformations and relations could be explored.

\section{Examples}

In this section, we illustrate \nrexamples{} realizations of \aname{} techniques on well-scoped examples.

\paragraph{Example 1: Infix, prefix, and postfix printing}

Abstract syntax trees (ASTs) are a common way to represent programs.
For example, an arithmetic expression \lstinline{(a + 3) * 2} could be represented as a tree \lstinline{Operation('*', Operation('+', Variable('a'), Constant(3)), Constant(2))}, where \lstinline{Operation} is the constructor for a binary operation node expecting the operator and the two operands as well as \lstinline{Constant} and \lstinline{Variable} the constructors for integer constants and named variables nodes.
We assume that we want to test a method \lstinline{as_string} that is implemented by every node.

\nparindent{}
The most common way to print an AST is assuming \emph{infix notation}, where, for a binary operation, the operator is printed between its operands.
As in the example above, a drawback of the infix notation is that operations need to be parenthesized when an outer operator has a higher precedence than an inner one.
As Listing~\ref{lst:asstring} shows, in our implementation, we account for this by explicitly checking whether the current operation is a multiplication and one of the child nodes an addition, in which case the addition needs to be parenthesized.

\nparindent{}
Implementations for the postfix and prefix notations are more compact and less error-prone.
The reason for this is that for these notations, the order is unambiguous.
For example, the expression \lstinline{(a + 3) * 2} would be printed as \lstinline{* + a 3 2} in prefix notation and as \lstinline{a 3 + 2 *} in postfix notation.

\nparindent{}
We realize our testing approach based on the insight that the original program can be modified by adding the prefix and postfix printing functions, as demonstrated by the functions \lstinline{as_string_prefix} and \lstinline{as_string_postfix} (Listing~\ref{lst:asstring}), which are more likely to be correct.
As a specific test oracle, we can test whether the same operations, variables, and constants are printed by deriving all three representations and comparing whether their individual tokens are the same, after removing the parentheses from the infix notation (see Listing~\ref{lst:orderstest}).
This allows detecting bugs where, for example, a mistake was made when assigning the parenthesized expression (\eg{}, \lstinline{right = '(' + left + ')'}).

\nparindent{}
This example shows how \aname{} can be useful to test a system where a complex component can be replaced with a simpler one, and the output of the system differs in a known way between the two components.
We implemented the test harness as an infinite loop.
In practice, it could be run for fixed inputs or a limited amount of time.

\begin{figure}
\begin{lstlisting}[label=lst:asstring, caption={Infix \lstinline{as_string} printing function as well as subsequently added prefix and postfix versions for testing. The line prefixed by \lstinline{+} were added to realize the \aname{} approach.},escapechar=!]
class Operation:
	def as_string(self):
		left = self.left.as_string()
		right = self.right.as_string()
		if self.operator == '*':
			if isinstance(self.left, Operation) and self.left.operator == '+':
				left = '(' + left + ')'
			if isinstance(self.right, Operation) and self.right.operator == '+':
				right = '(' + right + ')'
		return '%s %s %s' % (left, self.operator, right)

+	def as_string_prefix(self):
+		return '%s %s %s' % (self.operator, self.left.as_string_prefix(), self.right.as_string_prefix())
+
+	def as_string_postfix(self):
+		return '%s %s %s' % (self.left.as_string_postfix(), self.right.as_string_postfix(), self.operator)
\end{lstlisting}
\end{figure}

\begin{figure}
\begin{lstlisting}[label=lst:orderstest, caption={Test harness to validate that for each string representation, the same tokens are printed.}]
while True:
	tree = random_tree()
	tree_str = tree.as_string()
	tree_prefix_str = tree.as_string_prefix()
	tree_postfix_str = tree.as_string_postfix()
	infix_tokens = sorted(tree_str.replace('(', '').replace(')', '').split(' '))
	prefix_tokens = sorted(tree_prefix_str.split(' '))
	postfix_tokens = sorted(tree_postfix_str.split(' '))
	assert infix_tokens == prefix_tokens and infix_tokens == postfix_tokens
\end{lstlisting}
\end{figure}

\paragraph{Example 2: Monte Carlo Simulations}

\sloppy{}
Monte Carlo simulations are often used for simulating complex systems in physics~\cite{Hammersley2013}.
They are a class of algorithms that rely on repeated sampling to obtain numerical results.
Due to their non-deterministic nature, testing Monte Carlo simulations is generally difficult.
However, for this example, for simplicity, we assume that we would like to validate the correct implementation of a Monte Carlo simulation to estimate the value of pi, for which we know the ground truth.

\begin{figure}
\begin{lstlisting}[label=lst:piapprox, caption={Monte Carlo approximation of pi. The lines prefixed by \lstinline{+} were added to realize the \aname{} approach, while the lines prefixed by \lstinline{-} were removed.}]
- def get_pi_approximation():
+ def get_pi_approximation(n):
	inside = 0
-	for _ in range(1000000):
+	for _ in range(n):
		x = random.random()
		y = random.random()
		if x**2+y**2 <= 1:
			inside += 1

-	pi = 4*inside/1000000
+	pi = 4*inside/n
	return pi
\end{lstlisting}
    \vspace{-3mm}
\end{figure}

\begin{figure}
\begin{lstlisting}[label=lst:piapproxharness, caption={Test harness for the Monte Carlo approximation.}]
while True:
	pi_diff_inacc = abs(get_pi_approximation(10)- math.pi)
	pi_diff_acc = abs(get_pi_approximation(1000000)- math.pi)
	assert pi_diff_inacc >= pi_diff_acc
\end{lstlisting}
    \vspace{-5mm}

\end{figure}

\begin{figure}[tb]
    \includegraphics[width=0.8\columnwidth]{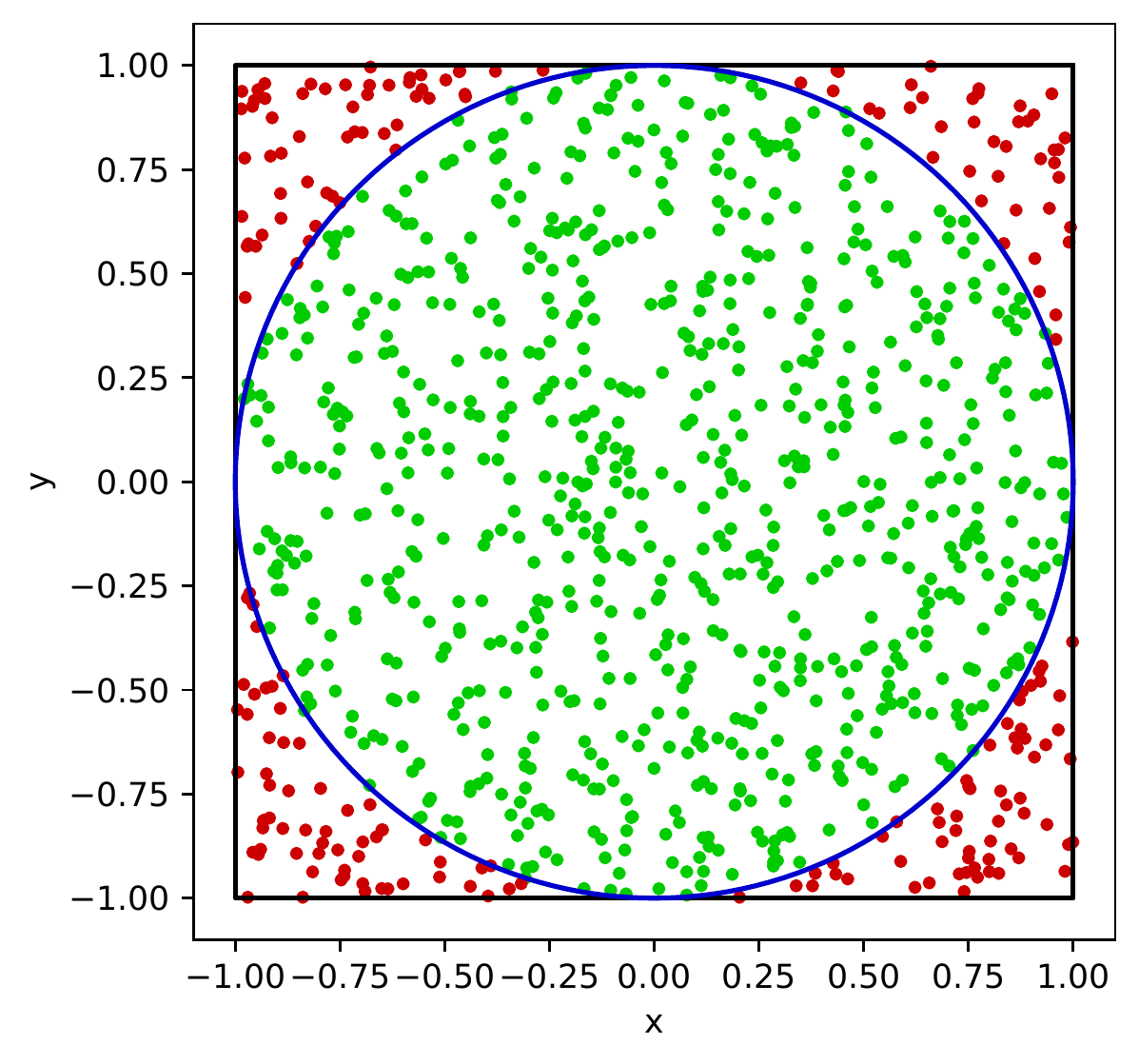}
    \caption[]{Visualization of the sampled points by the Monte Carlo simulation to compute pi.}
    \label{fig:montecarlopoints}
\end{figure}

\nparindent{}
Listing~\ref{lst:piapprox} shows how we can use Monte Carlo simulation to approximate the value of pi.
The initial \lstinline{get_pi_approximation} function expects no arguments and takes \lstinline{1,000,000} samples.
In each iteration, random x and y coordinates are drawn---the call to the random function returns a value from the interval $[0.0, 1.0)$, effectively only considering a rectangle around the circle.
We assume the circle's radius to be $1$; thus, checking \lstinline{x**2+y**2 <= 1} corresponds to checking whether the sample is part of the circle.
Figure~\ref{fig:montecarlopoints} illustrates the simulation; for the green points within the circle, the \lstinline{inside} variable is incremented.

\nparindent{}
The accuracy of the result of Monte Carlo simulations converges towards the real value given a large number of sampling steps~\cite{Graham2013}.
According to the law of large numbers, by sampling $n$ steps with $n \rightarrow \infty$, we expect to obtain an approximation of pi that is close to its real value.
As a practical realization of intramorphic testing, as shown in Listing~\ref{lst:piapproxharness}, we validate that we obtain a less accurate approximation of pi by obtaining a low number of samples ($10$ in the listing), rather than by using a large number of samples ($1,000,000$).
While there is no theoretical guarantee for this invariant to hold in theory, we observed that it holds in practice. 
To specify the number of samples, we modified \lstinline{get_pi_approximation} in Listing~\ref{lst:piapprox} to take take a parameter.

\begin{figure}[tb]
    \includegraphics[width=\columnwidth]{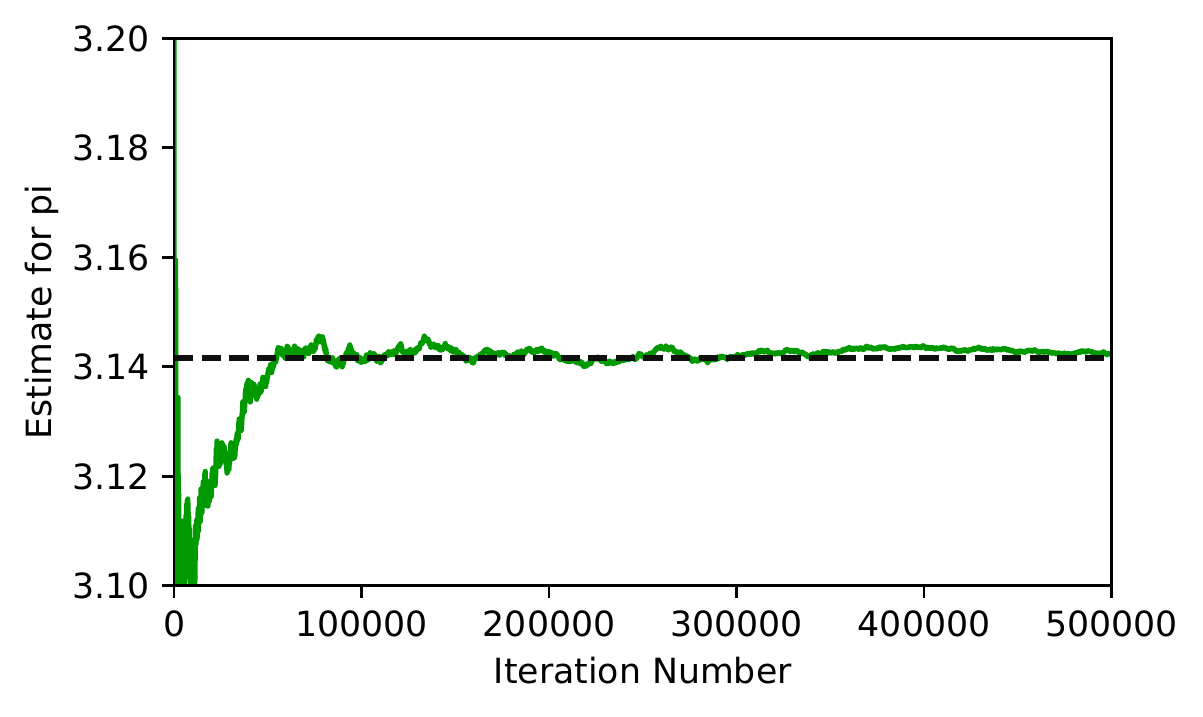}
    \caption[]{With an increasing number of samples, the approximated value of pi converges towards its real value.}
    \label{fig:montecarloestimate}
\end{figure}

\nparindent{}
This realization of an intramorphic testing approach demonstrates how numerical algorithms can be tested.
In this case, the program is modified to use a smaller (or larger) number of samples, with the expectation of obtaining a better estimate with more samples.
In practice, multiple samples could be taken to avoid false alarms caused by improbable cases where simulations with a fewer number of samples result in an approximation that is closer to the real value than simulations with a larger number of samples.
In contrast to the first example, rather than adding a function, we modified an existing function to take an additional parameter.

\paragraph{Example 3: Knapsack problem}

The knapsack problem is a well-studied combinatorics problem; many real-world combinatorics problems can be encoded as knapsack problems~\cite{Kellerer2004}. 
Given a knapsack with a given capacity, the goal is to pack items to maximize the value of items in the knapsack.
Each item has a value and weight associated with it.
The items' combined weight must not exceed the knapsack's capacity.
Various variations of the problems exist.
In this example, we consider the \emph{unbounded} knapsack problem, which places no restriction on the number of copies of each item.

\nparindent{}
Let us assume that our implementation solves the knapsack problem using a greedy approach, as illustrated in Listing~\ref{lst:greedyknapsack}, as computing an optimal solution could be considered too resource-intensive.
The parameter \lstinline{objects} is a list of triples \lstinline{((name, value, weight))}, that is, the name of the object, its value (sometimes referred to as \emph{profit}), and weight.
The second parameter~\lstinline{capacity} denotes the capacity of the knapsack.
The algorithm first sorts all items by \lstinline{value / weight}, and then adds items to the knapsack as long as its capacity is not exceeded.
The algorithm has no guarantees of computing an optimal solution.

\nparindent{}
The idea to test the implementation using intramorphic testing is that we can replace the greedy algorithm with the result of an algorithm that computes the optimal solution, knowing that the result should be as least as good as for the greedy algorithm.
The test harness is shown in Listing~\ref{lst:knapsackharness}.
The exhaustive algorithm that recursively explores all feasible solutions is implemented by the function \lstinline{knapsack_exhaustive} in Listing~\ref{lst:greedyknapsack}.
The core idea of the algorithm is that for every item at index \lstinline{item_index}, the algorithm explores separate branches assuming that the item is and is not included in the knapsack.

\nparindent{}
While we chose the knapsack algorithm as an example, we believe that intramorphic testing can be used for a wide range of algorithms used to solve NP-complete problems or provide approximate solutions; for example, register allocation is a NP-complete problem and the same idea could be used to compare a greedy linear-scan register allocator~\cite{Poletto1999} with a graph coloring one~\cite{Chaitin1981}.
In contrast to the other two examples, the change to the program was larger and more complex; however, the run-time characteristics and guarantees could be easier understood for such an optimal algorithm than for a greedy approach. 

\begin{figure}
\begin{lstlisting}[label=lst:greedyknapsack, caption={Greedy and optimal algorithms to solve the knapsack problem.}]
def knapsack_greedy(objects, capacity):
    packed = []
    cum_value = 0
    cum_weight = 0
    objects.sort(key=lambda triple : float(triple[1]) / triple[2], reverse=True)
    for (item, value, weight) in objects:
        while cum_weight + weight <= capacity:
            cum_weight += weight
            cum_value += value
            packed.append(item)
    return (packed, cum_value, cum_weight)

+ def knapsack_exhaustive(objects, capacity):
+    return knapsack_recursive(objects, capacity, 0, [], 0, 0)
+ def knapsack_recursive(objects, capacity, item_index, packed, cum_value, cum_weight):
+    if capacity <= 0 or item_index >= len(objects):
+        return (packed, cum_value, cum_weight)
+    current_included_profit, current_excluded_profit = 0, 0
+    space = False
+    cur_name, cur_value, cur_weight = objects[item_index]
+    if cur_weight <= capacity:
+        included_packed = packed.copy()
+        included_packed.append(cur_name)
+        current_included = knapsack_recursive(objects, capacity - cur_weight, item_index, included_packed, cum_value + cur_value, cum_weight + cur_weight)
+        space = True
+    current_excluded = knapsack_recursive(objects, capacity, item_index + 1, packed, cum_value, cum_weight)
+    if space and (current_included[1] > current_excluded[1]):
+        return current_included
+    else:
+        return current_excluded
\end{lstlisting}
\end{figure}

\begin{figure}
\begin{lstlisting}[label=lst:knapsackharness, caption={Comparing the results of the optimal with the greedy algorithm.}]
while True:
    capacity = random_capacity()
    vals = random_items()
    (_, val_exh, _) = knapsack_exhaustive(vals, capacity)
    print(knapsack_exhaustive(vals, capacity))
    (_, val_greedy, _) = knapsack_greedy(vals, capacity)
    assert val_exh >= val_greedy
\end{lstlisting}
\end{figure}

\changed{
\section{Discussion}

\paragraph{Examples}
We have presented \nrexamples{} realizations of \aname{} approaches on diverse, narrowly-scoped problems.
These examples demonstrate the approach's general idea as well as the challenges of designing intramorphic transformations.
We believe that in the future, various \aname{} techniques will be proposed that will operate at various levels and based on different insights, which could be, for example, specific to the domain or application.

\paragraph{Scope of the paper}
We have illustrated the approach's general idea on examples, and refrained from discussing and evaluating \aname{} techniques on large real-world applications.
We took inspiration from the original technical report on metamorphic testing, which was organized in a similar way; its practical merit was demonstrated in many innovative follow-up works~\cite{Segura2016,Chen2018}.

\paragraph{Cost}
Besides the potential benefit in finding bugs, \aname{} incurs both immediate and long-term costs.
The main immediate cost is the manual effort needed to implement the approach.
In addition, more code incurs a higher complexity; additional parameters and conditionals introduced may cause bugs.
Moreover, \aname{} also has a long-term cost, as the intramorphic transformations need to be maintained---changes in the codebase may require changes to the transformations.

\paragraph{Future research}
We believe that future research might lower the cost of \aname{} and make it more practical.
For example, rather than manually writing tests, future techniques could automatically suggest intramorphic relations.
As another example, similar to inline tests~\cite{LiuETAL22InlineTests}, approaches could be developed that facilitate co-evolvement of the source code and intramorphic tests.

}
\section{Conclusion}

We have presented \aname{}, \changed{a general approach} to tackling the test oracle problem, and illustrated it with various examples. 
The core idea of \aname{} is to modify a component of the system under test, anticipating a change on the program level.
If this anticipated change does not hold, we have discovered a bug in the system.
We believe that this technique will be widely useful to test systems while incorporating the domain knowledge of developers.

\begin{acks}
This research was supported by a Ministry of Education (MOE) Academic Research Fund (AcRF) Tier 1 grant.
\end{acks}

\bibliographystyle{ACM-Reference-Format}
{\balance
\bibliography{main}
}

\end{document}